\documentclass{aa}

\usepackage{graphicx}

\newcommand{\1}{{\Omega_M }}
\newcommand{\4}{{\Omega_X}}
\newcommand{\5}{{w_X}}

\newcommand{\beq}{\begin{equation}}
\newcommand{\eeq}{\end{equation}}

\begin{document}

\title{Probing the dark energy with strong lensing by clusters of galaxies}

\author{Mauro Sereno \inst{1} \fnmsep \inst{2}}

\offprints{M. Sereno, \\
\email{sereno@na.infn.it}}

\institute{Dipartimento di Scienze Fisiche, Universit\`{a} degli Studi di Napoli
``Federico II", Via Cinthia, Compl. Univ. di Monte S. Angelo, 80126
Napoli, Italia
\and Istituto Nazionale di Fisica Nucleare, Sez. di Napoli, Via Cinthia, Compl.
      Univ. di Monte S. Angelo, 80126 Napoli, Italia}

\date{Received 19 june 2001/ Accepted 27 June 2002}

\titlerunning{Probing dark energy}
\authorrunning{M. Sereno}

\abstract{Observations of clusters of galaxies that gravitationally lens faint
background galaxies can probe the amount and the equation of state,
$\5$, of the dark energy (quintessence) in the universe. Provided that
the mass profile and the mass normalization of the cluster are
determined, it is possible to constrain the cosmological parameters
that enter the lensing equations by means of the angular diameter
distances, by locating (either by observations of giant arcs and
magnification bias effect) the critical lines corresponding to known
redshift source populations of galaxies. This method can help to
distinguish between accelerating and decelerating models of the
universe. Furthermore, since the position of critical lines is
affected, especially in low-matter density universes, by the
properties of quintessence, the observations of a suitable number of
lensing clusters at intermediate redshifts can determine the equation
of state. A very preliminary application of the method to the cluster
CL~0024+1654 seems to support a flat accelerating universe dominated
by dark energy.
\keywords{cosmology: theory -- dark matter -- gravitational lensing -- galaxies:
clusters: individual: CL 0024+1654}}

\maketitle

\section{Introduction}
Observational cosmology has devoted large efforts in the last years to
characterize the energy content of the universe. Galaxy clustering
(Bachall \& Fan \cite{ba+fa98}; Carlberg et al. \cite{car&al98}) and
large-scale structure (Peacock et al. \cite{pe&al01}; Verde et al.
\cite{ver+al01}) observations favour models of a universe with a
subcritical matter energy density $\Omega_M$ (Turner \cite{tur00}).
Since, according to balloon-based measurements of the anisotropy of
the Cosmic Microwave Background Radiation (de Bernardis et al.
\cite{deb&al00}; Balbi et al. \cite{bal+al00}), the total of energy
content of the universe nearly equals the critical density (Jaffe et
al. \cite{ja&al00}; Pryke et al. \cite{pr+al02}), we expect that about
$2/3$ of the critical density is in form of dark energy (also called
quintessence). Furthermore, evidence coming from type Ia supernovae
that the universe is accelerating its expansion (Riess et al.
\cite{ri&al98}; Perlmutter et al. \cite{pe&al99}) demands a strongly
negative pressure for the dark energy ($w_X \equiv p_X
/\rho_X <
-1/3$, where $p_X$ and $\rho_X$ are, respectively, the pressure and
energy density of the dark energy). These observations, together with
other constraints coming from the age of the universe, gravitational
lensing statistics and Ly$\alpha$ forest, support a geometrically flat
universe (Harun-or-Rashid \& Roos \cite{ha&ro01})($\Omega_M +
\Omega_X =1$, where $\Omega_X$ is the dark energy density parameter of the
universe) with $\Omega_M \sim 0.3$-$0.4$ and a constant equation of
state $-1
\leq w_X
\stackrel{<}{\sim}
-0.4$ (Waga \& Miceli \cite{wa+mi98}; Wang et al. \cite{wan+al00}) at the $68\%$ confidence
level or better according to a concordance analysis (Wang et al.
\cite{wan+al00}). A less conservative maximum likelihood analysis
suggests a smaller range for the equation of state, $-1 \leq w_X
\stackrel{<}{\sim} -0.6$ (Perlmutter et al. \cite{pe&al99b}; Wang et
al. \cite{wan+al00}; Bean \& Melchiorri \cite{be+me02}). According to
these results, in what follows, without being explicitly stated, we
will assume a flat universe.

After the first proposal of dark energy (the cosmological constant,
$w_X=-1$), many other candidates have been suggested. One interesting
idea is that the energy density is provided by a scalar field rolling
down an almost flat potential (Caldwell et al. \cite{ca&al98}; Ratra
\& Peebles \cite{ra&pe98}; de Ritis et al. \cite{rit&al00}; Rubano \&
Scudellaro \cite{ru&sc01}). Other possibilities are represented by a
fluid with a constant equation of state, called $X$-matter (Chiba et
al. \cite{ch&al97}; Turner \& White \cite{tu+wh97}), or by a network
of light non-intercommuting topological defects (Vilenkin
\cite{vil84}; Spergel \& Pen \cite{sp+pe97}) ($w_X =-m/3$ where $m$ is
the dimension of the defect: for a string, $m=1$; for a domain wall,
$m=2$). Generally, the equation of state $w_X$ evolves with the
redshift, and the feasibility of reconstructing its time evolution has
been investigated (Cooray
\& Huterer \cite{co&hu99}; Chiba \& Nakamura \cite{ch&na00};
Saini et al. \cite{sa&al00}; Goliath et al. \cite{gol&al01}; Huterer
\& Turner \cite{hu&tu01}; Maor et al. \cite{ma&al00};
Nakamura \& Chiba \cite{na&ch01}; Wang \& Garnavich \cite{wa+ga01};
Yamamoto \& Futamase \cite{ya+fu01}; Corasaniti \& Copeland
\cite{co+co02}). Since in flat Friedmann-Lema\^{\i}tre-Robertson-Walker
(FLRW) models the distance depends on $w_X$ only through a triple
integral on the redshift (Maor et al. \cite{ma&al00}), $w_X (z)$ can
be determined only given a prior knowledge of the matter density of
the universe (Goliath et al. \cite{gol&al01}; Weller \& Albrecht
\cite{we+al01a}; Gerke
\& Efstathiou \cite{ge+ef02}). In what follows, we will consider only the case of a
constant equation of state.

Although the listed results are really compelling, it is still useful
to develop new tools for the determination of the cosmological
parameters. Many of the discussed methods are affected by
shortcomings, like poorly controlled systematic errors or large
numbers of model parameters involved in the analysis. An independent
constraint can improve the statistical significance of the statement
about the geometry of the universe and can disentangle the degeneracy
in the space of the cosmological parameters.

Gravitational lensing systems have been investigated as probes of dark
energy. Gravitational lensing statistics (Waga
\& Miceli \cite{wa+mi98}; Cooray
\& Huterer \cite{co&hu99}; Wang et al. \cite{wan+al00}; Zhu \cite{zhu00}),
effects of large-scale structure growth in weak lensing surveys
(Benabed
\& Bernardeau \cite{be+be01}) and Einstein rings in galaxy-quasar systems
(Futamase \& Yoshida \cite{fu&yo00}; Yamamoto \& Futamase
\cite{ya+fu01}) are very promising ways to test quintessence. Here, we
propose to investigate clusters of galaxies acting as lenses on
background high redshift galaxies. The feasibility of these systems to
provide information on the universe is already known (Paczy\'{n}ski
\& Gorski \cite{pa&go81}; Breimer \& Sanders \cite{br+sa92}; Fort et
al. \cite{for+al97}; Link \& Pierce \cite{li&pi98}; Lombardi \& Bertin
\cite{lo&be99}; Gautret et al. \cite{ga&al00}). Provided that the
modeling of the lens is constrained, once both arc positions and its
redshift are measured, it is possible to gain an insight into
second-order cosmological parameters contained in angular diameter
distances ratios (Chiba \& Takahashi \cite{ch+ta01}; Golse et al.
\cite{go&al01}). In addition to observations of arcs, a statistical
approach based on magnification bias (Broadhurst et al.
\cite{bro+al95}; Fort et al. \cite{for+al97}; Mayen
\& Soucail \cite{ma&so00}) can as well locate the critical lines (locations of maximum amplification)
corresponding to background source populations.

In this paper, we will explore the feasibility of clusters of galaxies
acting as lenses in probing both the amount and the equation of state
of quintessence in the universe, assumed to be flat. In Sect.~2, we
outline the method. Section~3 is devoted to an application to the
cluster of galaxies CL 0024+1654. In Sect.~4, we discuss some
systematics affecting the method. Some final considerations are
presented in Sect.~5.

\section{How critical lines depend on dark energy}

\begin{figure}
   \resizebox{\hsize}{!}{\includegraphics{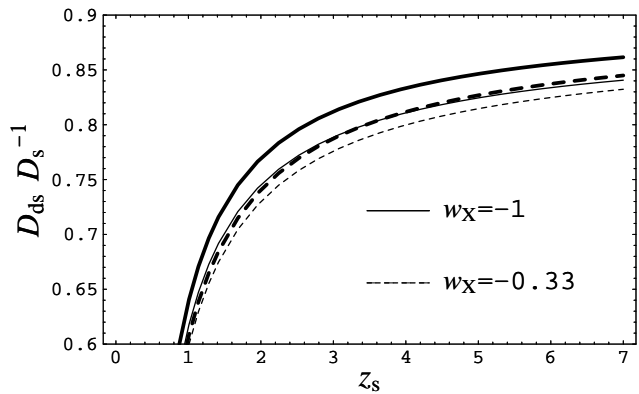}}
   \caption{The ratio of distances $D_{ds}/D_s$ as a function of the source
   redshift for a deflector at $z_d =0.3$, for different sets of cosmological
   parameters. The thick lines
   correspond to $\1=0.3$; the thin lines to $\1=0.5$. The full and dashed
   lines correspond to, respectively, $w_X=-1$ and $w_X=-1/3$.}
   \label{ratio_zs}
 \end{figure}

\begin{figure}
   \resizebox{\hsize}{!}{\includegraphics{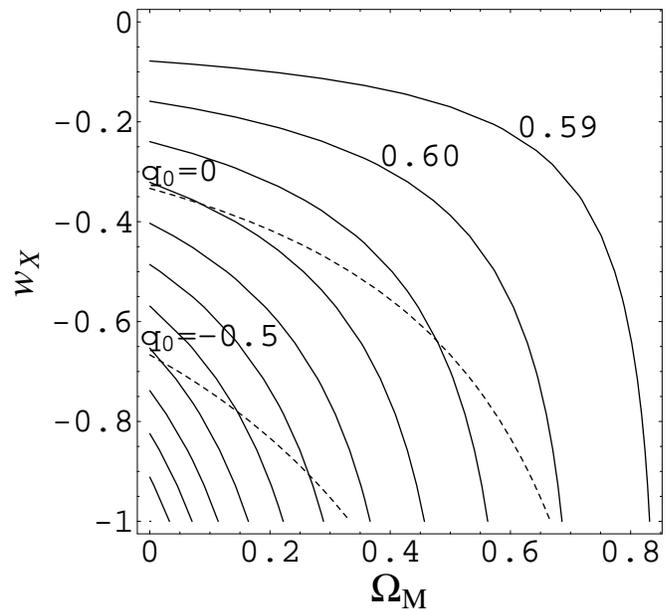}}
   \caption{Contours of equal $D_{ds}/D_s$ on the $(\Omega_M,w_X)$
   plane for $z_d=0.3$ and $z_s =1$.
   Each contour is drawn with a step of 0.01. The value of the contours
   increases from the top right corner to the bottom left corner.
   The thin dashed lines correspond to lines of constant $q_0$.}
   \label{omega_M-w_X}
 \end{figure}

\begin{figure}
   \resizebox{\hsize}{!}{\includegraphics{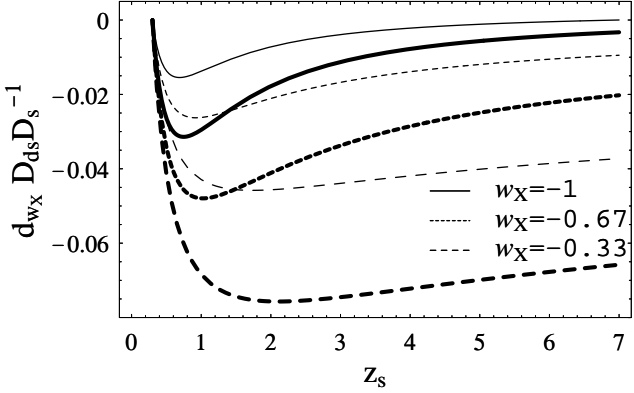}}
   \caption{The derivative of the ratio of distances $D_{ds}/D_s$ with
   respect to $w_X$ for a lens at $z_d =0.3$ as a function of the source
   redshift, for different values of the equation of state. The full lines
   correspond to $w_X=-1$; the dashed lines to $w_X=-2/3$; the long-dashed
   lines to $w_X=-1/3$. The thick (thin) lines are for $\Omega_M=0.3(0.5)$.}
   \label{der-cri-1}
 \end{figure}

\begin{figure}
   \resizebox{\hsize}{!}{\includegraphics{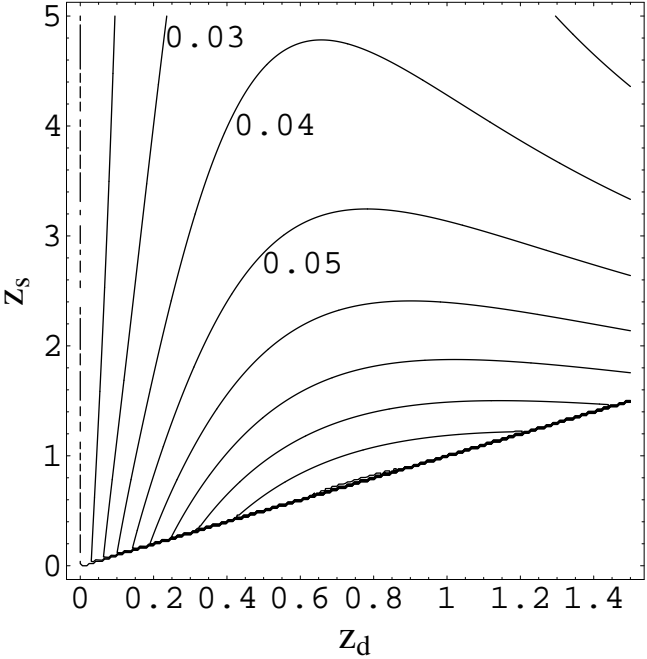}}
   \caption{The relative variation between the ratio of distances $D_{ds}/D_s$
   for two cosmological models, ($\1=0.3,w_X=-1$) and ($\1=0.3,w_X=-1/3$), in the
   $(z_d,z_s)$ plane. Each contour is drawn with a step of $0.01$.}
   \label{zd-zs}
 \end{figure}

The study of critical lines in a gravitational lensing system is a
potentially important tool to probe the content of dark energy in the
universe and to constrain its equation of state, as already shown in
the case of galaxy-quasar lensing in Futamase \& Yoshida
(\cite{fu&yo00}) and Yamamoto \& Futamase (\cite{ya+fu01}). This type
of cosmological investigations requires an accurate modeling of the
lens, the observation of a critical line and the knowledge of the
redshifts of both the lens and the deflected source (Breimer
\& Sanders \cite{br+sa92}; Fort et al. \cite{for+al97}; Link \& Pierce \cite{li&pi98}).

As an example for our quantitative considerations, let us consider as
deflecting cluster a singular isothermal sphere (SIS). The projected
density mass $\Sigma$ of the SIS is
\begin{equation}
\label{cri1}
\Sigma (\theta)=\frac{\sigma^2}{2G}\frac{1}{D_d\theta},
\end{equation}
where $G$ is the Newtonian constant of gravitation, $\sigma$ the
velocity dispersion, $D_d$ the angular diameter distance to the
deflector and $\theta$ the angular position in the sky. For a
spherically symmetric lens, the tangential critical line is determined
by (Schneider et al. \cite{sef})
\begin{equation}
\label{cri1bis}
\theta_t=\sqrt{ \frac{4 G M(\theta_t)}{c^2} \frac{D_{ds}}{D_d D_s}},
\end{equation}
where $c$ is the velocity of the light and $M(\theta)$ is the lens
mass within the radius $\theta$; $D_{ds}$ and $D_s$ are the angular
diameter distances to the source from, respectively, the lens and the
observer. For the SIS, Eq.~(\ref{cri1bis}) reduces to
\begin{equation}
\label{cri2}
\theta_t=4 \pi \left( \frac{\sigma}{c}\right)^2 \frac{D_{ds}}{D_s}.
\end{equation}
Once $\theta_t$ and $\sigma$ are known, the ratio of distances
$D_{ds}/D_s$ can be determined.

The dependence on the cosmological parameters is contained in the
angular diameter distance. In a flat FLRW universe, the angular
diameter distance between an observer at $z_d$ and a source at $z_s$
is
\begin{eqnarray}
\label{cri3}
\lefteqn{D(z_d, z_s)=\frac{c}{H_0}\frac{1}{1+z_s}} \\
& &  {\times}\int_{z_d}^{z_s} \frac{dz}{\sqrt{\Omega_M
(1+z)^3+(1-\Omega_M)(1+z)^{3(w_X+1)}}}; \nonumber
\end{eqnarray}
$H_0$ is the today Hubble parameter. At high redshift, the
pressureless matter density overcomes the dark energy; for large $z_d$
and small $w_X$, $D_{ds}$ is nearly insensitive to the equation of
state. For the expression of the distance in inhomogeneous universes,
we refer to Sereno et al. (\cite{ser+al01}, \cite{ser+al02}).

Let us go, now, to examine the feasibility of determining $\5$ with
observations of strong lensing events in clusters of galaxies by the
study of the ratio of distances $D_{ds}/D_s$. Once the lens redshift
is fixed, $D_{ds}/D_s$ first increases rapidly with the source
redshift and, then, for $z_s$ greater than $2.5$, is nearly constant
(Asada \cite{asa97}; Fort et al. \cite{for+al97}), as can be seen in
Fig.~(\ref{ratio_zs}). The change with the cosmological parameters can
be significant. The ratio increases with decreasing $\1$ and with dark
energy with large negative pressure, i.e. it is maximum in the case of
the cosmological constant. The variations with $\1$ and $\5$ are
comparable. Changing $\1$ from $0.3$ to $0.5$ has the same effect of
increasing $\5$ from $-1$ to $-1/3$, so that $D_{ds}/D_s$ is nearly
indistinguishable in a universe with $\1=0.3$ filled in with string
networks and in a model with $\1 =0.5$ and cosmological constant.

To quantify the dependence of $D_{ds}/D_s$ with the cosmological
parameters, we consider fixed redshifts for the lens and the source,
see Fig.~(\ref{omega_M-w_X}). The ratio is quite sensitive to $\1$.
The variations due to changes in $\1$ for $w_X=const.$ are greater
than in the case of the constant deceleration parameter $q_0
\equiv (1+3w_X(1-\1))/2$. For $z_d=0.3$, $z_s=1$, and $\1$ ranging
from $0$ to $0.6$, when $q_0=0$ the variation is $\sim 4\%$; when
$w_X=-1$, the variation is $\sim 15\%$. The dependence on the
cosmological parameters is maximum for high negative values of $q_0$,
i.e. the region today preferred by observations. For some particular
pairs $(z_d,z_s)$, i.e. for low lens redshifts and sources very near
to the deflector, the ratio is nearly constant on lines of constant
deceleration parameters; these properties suggest that the method of
the critical line can help to distinguish between accelerating and
decelerating universes. The dependence of $D_{ds}/D_s$ on the equation
of state increases for low matter density universes and the
sensitivity nearly doubles for small changes in $\1$: for $z_d =0.3$
and $z_s =1$, the relative variation from $w_X
=-1$ to $\5 =-1/3$ is $1.9\%$ ($3.4\%$) when $\1 =0.5$ ($0.3$). The
sensitivity is maximum for intermediate $\5$; for large negative
pressure ($w_X \stackrel{<}{\sim} -0.9$), the ratio is nearly
independent of variations of the equation of state.

In Fig.~(\ref{der-cri-1}), the derivative of the ratio $D_{ds}/D_s$
with respect to $w_X$ is plotted as a function of the redshift of the
source once the redshift of the deflector is fixed. The derivative is
negative for a large range of redshifts of both source and deflector.
Transitions from negative to positive values occur for very negative
$w_X$. The source redshift where the derivative cancels out decreases
with increasing $\Omega_M$ and $z_d$: for $z_d=0.3$ ($0.6$),
$\Omega_M=0.3$ and $w_X=-1$, the derivative is null at $z_s \simeq
7.0$ ($2.1$). The sign of the derivative determines, when the equation
of state changes, as the angular position of the critical lines moves:
when the derivative is negative (positive), as the equation of state
increases (i.e. as $w_X$ moves from $-1$ to $0$), the angular radius
in the sky of the critical line, for fixed source and deflector
redshifts, decreases (increases).

The modulus of the derivative is an estimate of the dependence of the
ratio on $w_X$. Independently of the value of $z_d,\
\Omega_M$ and $w_X$, the dependence on $w_X$ first increases and takes its
maximum at an intermediate source redshift, and then decreases quite
slowly. For dark energy in the form of a cosmological constant
($w_X=-1$), $z_d=0.3$ and $\Omega_M =0.3$, the maximum is at $z_s \sim
0.75$. For increasing $w_X$, the maximum moves to higher redshifts:
for domain walls ($w_X =-2/3$), the maximum is at $z_s \sim 1.02$.
From Fig.~(\ref{der-cri-1}), we see that for a large range of $w_X$
and $\Omega_M$ the maximum is at $z_s \stackrel{<}{\sim} 2$. This
trend of the derivative is connected to the properties of the ratio
$D_{ds}/D_s$, that flattens at higher source redshifts.

Now, we want to search for the optimal lens and source configuration
in order to discriminate among quintessence models. For illustration,
we choose two universes with the same content of matter ($\1 =0.3$)
but different $\5$; we consider a cosmological constant ($\5=-1$) and
string networks ($w_X =-1/3)$. In Fig.~(\ref{zd-zs}), we scan the
$(z_d,z_s)$ plane plotting the relative variation between the two
pairs of cosmological parameters. For a given lens redshift, the best
$z_s$ is very close to the deflector, i.e. a couple of redshifts
corresponding to the rising part of the ratio $D_{ds}/D_s$; the
sensitivity decreases for larger and larger source redshifts. So, the
configurations with high sensitivity to the quintessence are those
with very low cross section for strong lensing events. On the other
hand, given a background population at $z_s
\stackrel{>}{\sim} 1$, the optimal lens is a quite high redshift
cluster at $z_d \sim 0.7$; however, the dependence on the quintessence
is nearly constant for lenses at $z_d \stackrel{>}{\sim} 0.6$

In order to estimate the accuracy of the determination of the equation
of state, the variation induced on $D_{ds}/D_s$ by $w_X$ must be
compared to the error within which the parameters of the lens are
known. For the SIS, the error in the estimate of the ratio of
distances is
\begin{equation}
\label{cri4}
\left| \Delta \left( \frac{D_{ds}}{D_s} \right) \right|=\sqrt{
4\left| \frac{\Delta \sigma}{\sigma} \right|^2 + \left| \frac{\Delta
\theta_t}{\theta_t} \right|^2} \left|
\frac{D_{ds}}{D_s}\right|,
\end{equation}
where $\Delta \sigma$ and $\Delta \theta_t$ are the errors,
respectively, on the velocity dispersion and $\theta_t$. Usually, the
largest uncertainty in the modeling of a lens comes from the error in
the measurement of the velocity dispersion. Catalogues of galaxy
velocities in lensing clusters are of the order of $50$, so that the
uncertainty on $\sigma$ is $\sim 15\%$. $\Delta
\theta_t$ comes from the accuracy of the location of the arc and its
radial thickness and from the uncertainty on the geometrical
properties of the lens, i.e. the accuracy of the location of the
center, typically chosen to coincide with the brightest cluster
galaxy, and the ellipticity of the mass distribution. For tangential
arcs at $\theta_t \sim 20 \arcsec$ (Williams et al. \cite{wil+al99}),
an error as large as $\sim 1 \arcsec$ can contribute a $5 \%$ error.
The error on $\theta_t$ is generally negligible with respect to the
error in the mass normalization and will not be considered in the rest
of this section. The variation on $D_{ds}/D_s$ connected to changes in
the equation of state can be expressed as
\begin{equation}
\label{cri5}
\left| \frac{\partial}{\partial w_X} \left( \frac{D_{ds}}{D_s} \right) \right|\Delta w_X ,
\end{equation}
and so, comparing Eq.~(\ref{cri4}) and Eq.~(\ref{cri5}), for $N$
clusters we have a statistical error of
\begin{eqnarray}
\label{cri6}
\lefteqn{
\Delta w_X \stackrel{>}{\sim}\frac{2}{\sqrt{N}}\left|
\frac{\Delta \sigma}{\sigma} \right| \left \langle \left|  \frac{D_{ds}}{D_s}\right|\left|
\frac{\partial}{\partial w_X} \left( \frac{D_{ds}}{D_s} \right)
\right|^{-1} \right \rangle} \nonumber \\
& & = \frac{2}{\sqrt{N}}\left|
\frac{\Delta \sigma}{\sigma} \right| \left \langle \left|
\frac{\partial}{\partial w_X} \ln \left( \frac{D_{ds}}{D_s} \right) \right|^{-1}
\right \rangle ,
\end{eqnarray}
where the average is on the redshifts of the critical lines. The error
in the determination of $w_X$ increases with $\Delta \sigma$ and
decreases with the derivative. Since the error induced by the velocity
dispersion is proportional to the ratio of distances $D_{ds}/D_s$, see
Eq.~(\ref{cri4}), and the variation induced by $w_X$ is proportional
to the derivative, see Eq.~(\ref{cri5}), the uncertainty in the
estimate of $w_X$ is inversely proportional to the logarithmic
derivative of $D_{ds}/D_s$, i.e. to the relative variation of
$D_{ds}/D_s$. The properties of the logarithmic derivative with
respect to the cosmological parameters $\Omega_M$ and $w_X$ are the
same of the ordinary derivative; the main difference is the
disappearance of the minimum. As we have seen before, the uncertainty
in the equation of state, given a deflector redshift, increases with
$z_s$ and decreases for quintessence with $w_X$ far away from $-1$.

The case of the cosmological constant is the more problematic one
since the derivative can cancel out (when $\1=0.3$ and $z_s=1.5 $, the
derivative is null at $z_d \simeq 1.13$). However, clusters at
intermediate redshift $(z_d \sim 0.4)$ are quite stable with respect
to the error in the equation of state.

As we shall see in the next section, it is possible to obtain
information from a single cluster of galaxies on more than one
critical line. So, using in Eq.~(\ref{cri6}) the number $N$ of
clusters, the lower limit on $\Delta w_X$ is overestimated. Given a
typical error of $\sim 15 \%$ on $\sigma$, we can use Eq.~(\ref{cri6})
to estimate the number of deflectors necessary for estimating $w_X$
within a given uncertainty. For mean redshifts of $\langle z_d \rangle
=0.4$ and $\langle z_s
\rangle =1.2$, an uncertainty of $\Delta w_X \simeq 0.25$ needs
$\sim 75$ ($\sim 120)$ lensing clusters in a universe with $\1 =0.3$
and $\5=-1/3$ ($-0.5$). $N$ increases with dark energy with large
negative pressure and large values of $\1$. As discussed, the method
is unable to constrain the equation of state in the extreme case of a
cosmological constant, when $\Delta w_X \simeq 0.25$ needs $\sim 800$
clusters and $\Delta w_X \simeq 0.5$ needs $\sim 200$ clusters. In
general, to distinguish dark energy with an intermediate value of
$w_X$ from a cosmological constant at $95\%$ confidence level, in a
low matter density universe, we need $100$-$200$ strong lensing
events. These simple estimates are in agreement with the results in
Yamamoto \& Futamase (\cite{ya+fu01}).

Together with spectroscopic analyses, X-ray observations of a lensing
cluster can help to estimate the absolute mass of the deflector. The
projected X-ray cluster mass, under the hypotheses of isothermal and
hydrostatic equilibrium, is proportional to the cluster gas
temperature, $T_X$, and $D_d$ (Wu \cite{wu00}): X-ray data alone
cannot determine the mass without a prior knowledge of cosmological
parameters. However, it has been shown that the relation between
$\sigma$ and $T_X$ is not affected by cosmic evolution and is
consistent with the isothermal scenario, $\sigma \propto T_X^{0.5}$
(Wu et al. \cite{wu+al98}). Once calibrated this relation, X-ray
observations obtained with the new generation of telescopes can
considerably enlarge the data sample of lensing clusters with known
mass and help to disentangle the effect of cosmology and mass
normalization of the deflector.

\section{CL~0024+1654}

Now, let us consider the application of the method outlined in Sect. 2
to a well studied cluster of galaxies, CL~0024+1654, in order to test
the feasibility of what we are proposing, and how good the results can
be.

\begin{figure}
   \resizebox{\hsize}{!}{\includegraphics{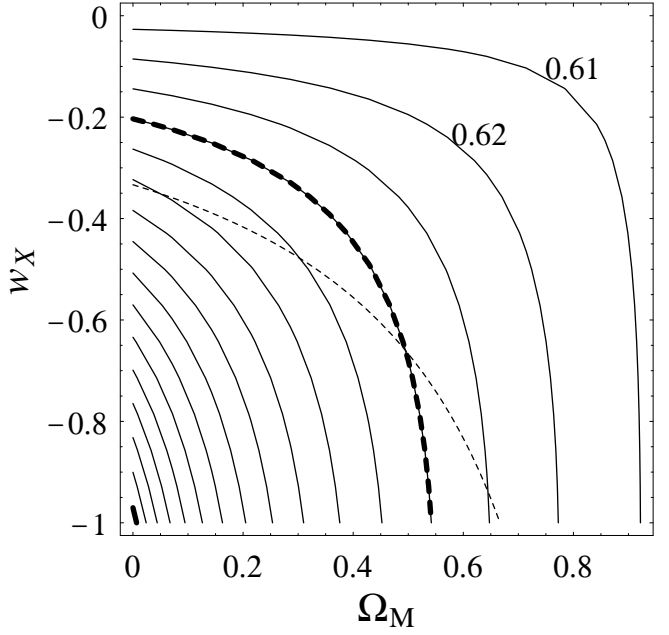}}
   \caption{Contours of equal $D_{ds}/D_s$ on the $(\Omega_M,w_X)$ plane
    for CL~0024+1654 ($z_d=0.395$) and its multiple arc ($z_s=1.675$).
    Each contour is drawn with a step of 0.01. The value of the contours
    increases form $0.61$ in the top right corner to $0.76$ in the
    lower left corner. The thick lines correspond to the data in Eq. (\ref{cl3}).
    The thick full line corresponds to the best parameters; the dashed
    one to the lower limit. The thin dashed line
    separates accelerating universes (below) from decelerating ones
    (above).}
   \label{omega_M-w_X_tys}
 \end{figure}

CL~0024+1654 is one of the best investigated lenses in the universe.
It is an optically rich cluster of galaxies, with a relaxed structure
without a single central dominant cluster galaxy, at $z=0.395$ and
with a velocity dispersion of $\sigma =1050  {\pm} 75$ km s$^{-1}$
(Dressler et al. \cite{dr&al99}; Czoske et al. \cite{czo+al01}; Czoske
et al. \cite{czo+al02}). This is the formal velocity dispersion
estimated with the assumptions of virial equilibrium and random galaxy
velocities, so that the reported error is a purely statistical one. We
will consider the effect of some possible systematics in the next
section. This value of $\sigma$ is consistent with lensing
observations (Shapiro \& Iliev \cite{sh&il00}). X-ray data (Soucail et
al. \cite{sou+al00}; B\"{o}hringer et al. \cite{boh+al00}) also support a
regular morphology with no significant substructures. The measured
value of $T_X
=5.7^{+4.9}_{-2.1}$ is compatible with the observed velocity
dispersion. A single background galaxy behind CL~0024+1654, at
spectroscopic redshift $z=1.675$ (Broadhurst et al. \cite{bro+al00}),
is imaged in a well known multiple arc at $\theta_t
= 30.5\arcsec$ (Kassiola et al. \cite{ka&al92}; Wallington et al.
\cite{wal+al95}, Smail et al. \cite{sma+al96}; Tyson et al.
\cite{tys+al98}). Images are characterized by a bright elongated knot,
surrounded by a low surface brightness halo. The knot comprises two
peaks, with separations ranging from $0.5\arcsec$ to $1.1\arcsec$,
roughly consistent with the relative lengths of the various arc
components (Smail et al. \cite{sma+al96}). Given this peculiar
morphology, we assume an indetermination on the critical radius
$\Delta
\theta_t \sim 0.7\arcsec$. We do not take into account the error on
the position of the centre; in the analyses considered here, it is
determined as a free parameter in the lensing reconstruction. Based on
deep images with the Hubble Space Telescope, Tyson et al.
(\cite{tys+al98}) performed a multi-parameter fit, including a number
of small deflecting ``mascons", to the mass profile. Each mascon was
parameterized with a power-law model (Schneider et al. \cite{sef}),
\begin{equation}
\label{cl1}
\Sigma (\theta ) \propto \frac{1+\gamma \left( \frac{\theta}{\theta_c}\right)^2}{\left[
1+\left(\frac{\theta}{\theta_c}\right)^2 \right]^{2-\gamma}}\ ,
\end{equation}
where $\theta_c$ is the core radius and $\gamma$ is the slope.
Remarkably, they found that more than $98\%$ of the cluster matter is
well represented by a single power-law model centred near the
brightest cluster galaxies with $\theta_c =10.0{\pm} 0.9\arcsec$ and
$\gamma =0.57 {\pm} 0.02$, slightly shallower than an isothermal sphere
($\gamma_{SIS}=0.5$). To disentangle the effect of cosmology and
absolute mass, we have to fix the central density $\Sigma_0$
independently of lensing data. It is (Bonnet et al. \cite{bon+al94}),
\begin{equation}
\label{cl1b}
\Sigma_0=\frac{8\alpha \sigma^2}{3 \pi G \theta_c} \frac{I^2_{(1+\alpha)/2}}{I_{\alpha}} \frac{1}{D_d},
\end{equation}
with $\alpha=2(1-\gamma)$ and
$I_\beta=\int_{0}^{\infty}(1+u^2)^{-\beta}du$. For a power-law model,
the angular position $\theta_t$ of the tangential critical line is
related to the angular diameter distances and the parameters of the
lens by
\begin{equation}
\label{cl2}
\left( \frac{\theta_t}{\theta_c}\right)^2=\left( \frac{4\pi G}{c^2}\Sigma_0
\frac{D_d D_{ds}}{D_s}\right)^{\frac{1}{1-\gamma}} -1.
\end{equation}
Substituting in Eq. (\ref{cl2}) for $\Sigma_0$ and using the fit
parameters, we get an estimate for the ratio $D_{ds}/{D_s}$. It is
\begin{equation}
\label{cl3}
\frac{D_{ds}}{D_s}(z_d=0.395,z_s=1.675)=0.76^{+0.18}_{-0.12}.
\end{equation}
 The main term in the error budget comes from the indetermination in
the velocity dispersion which contributes $\sim 75\%$ of the total
error. In Fig.~(\ref{omega_M-w_X_tys}), we show the dependence of $
D_{ds}/{D_s}$ on the cosmological parameters for a lens-source
configuration as in CL~0024+1654; the values of $(\Omega_M,w_X)$
compatible with the estimate in Eq. (\ref{cl3}) are also plotted. Low
matter density universes ($\1
\stackrel{<}{\sim} 0.55$), which are accelerating their expansion, are
favoured. We find $-1 \leq w_X
\stackrel{<}{\sim} -0.2$, with the lower values of $\Omega_M$
corresponding to the higher values of the equation of state.

\begin{figure}
        \resizebox{\hsize}{!}{\includegraphics{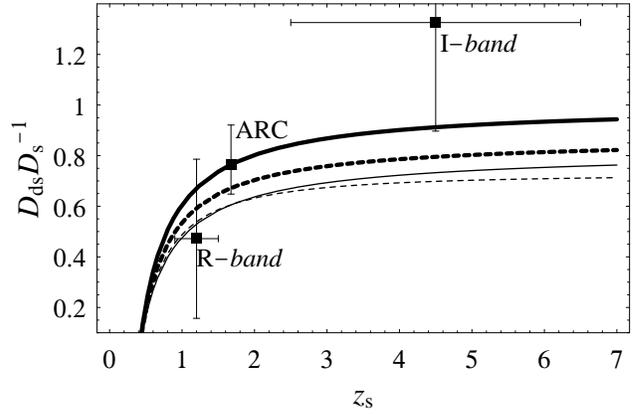}}
        \caption{$D_{ds}/D_s$ as a function of the source redshift for
        CL~0024+1654 ($z_d=0.395$) for different cosmological models.
        The thick lines are for flat models with quintessence. The full thick
        line for $\4 =1$  and $\5 =-1$ (de Sitter universe); the dashing thick line is for
        $\Omega_M =0.3$ and $w_X=-1$. The thin
        lines are for universes with pressureless matter alone. The full line
        is for an Einstein-de Sitter universe ($\1 =1$); the thin dashed
        line is for an open universe ($\Omega_M =0.1$). ``R-$band$" indicates the
        data derived from Dye et al. (\cite{dye+al02}) (the error in $\sigma$
        is not considered); ``ARC" is the data in Eq.~(\ref{cl3}); ``I-$band$" indicates the
        data from the depletion curve in Fort et al.~(\cite{for+al97}).}
        \label{tys+fmd+dye}
\end{figure}

The method of the depletion curves, i.e. the variation along the
radial direction in the surface density of background galaxies around
a massive cluster of galaxies, has been employed to further study
CL~0024+1654 (Fort et al. \cite{for+al97}; van Kampen \cite{van98};
R\"{o}gnvaldsson et al. \cite{ro&al01}; Dye et al. \cite{dye+al02}).
Observations of the magnification bias have been obtained in the $B$-
and $I$-band (Fort et al. \cite{for+al97}) and in the $U$- and
$R$-band (R\"{o}gnvaldsson et al. \cite{ro&al01};  Dye et al.
\cite{dye+al02}). Extrapolating Hubble Space Telescope data to their
detection limit, Dye et al. (\cite{dye+al02}) obtained, for the
background $R$-galaxies, a mean redshift of $\langle z_s \rangle =1.2
{\pm}0.3$. From a fit to the SIS profile of the depletion curve in the
$R$-band, the location of the critical curve comes out at $15{\pm}10
\arcsec$ (Dye et al. \cite{dye+al02}). Using these estimates in
Eq.~(\ref{cri2}), we can obtain a second constraint on the ratio
$D_{ds}/D_s$; unfortunately, as can be seen in
Fig.~(\ref{tys+fmd+dye}), the uncertainties completely hide the second
order effect of the cosmological parameters on the ratio of distances
$D_{ds}/D_s$.

A more interesting result can be obtained from the $I$-band. As
discussed in Fort et al. (\cite{for+al97}), the angular radius where
the depletion curve starts to increase locates the last critical line,
that is the critical line corresponding to the farther source
population. The last critical line at $\sim 60\arcsec$ in the $I$-band
corresponds to background galaxies at redshift $2.5< z < 6.5$;
however, about $20\%$ of the very faint $I$ selected galaxies should
be above $z=4$. As noted in van Kampen (\cite{van98}), given the very
low density of the background $I$-galaxies, an appropriate radial
binning to study the radial profile of the magnification bias is $30
\arcsec$. So, we will consider an error of $15 \arcsec$. This estimate of
the location of the last critical line is independent of the assumed
mass profile, and can be used in Eq.~(\ref{cl2}) to obtain a new
constraint on $D_{ds}/D_s$, see Fig.~(\ref{tys+fmd+dye}). Since
$D_{ds}/D_s$ is nearly flat for $z_s \stackrel{>}{\sim} 2.5$, the
value of the ratio of distances is quite insensitive to the value of
$z_s$ corresponding to the last critical line.

Some interesting considerations are obtained from the variation of the
ratio $D_{ds}/D_s$ with the redshift of the source. Figure
(\ref{tys+fmd+dye}) shows $D_{ds}/D_s$ for a lens at $z=0.395$ and for
various cosmological models. Models without dark energy are rejected,
with no regard to the value of the pressureless matter density: both
open ($\Omega_M <1$) and flat (the Einstein-de Sitter model, with
$\Omega_M
=1$) dark matter models are very poorly consistent with the
experimental points. On the other hand, flat universes with
quintessence are in agreement with the data. In particular, the data
from the $I$-band analysis are marginally compatible (at the $68\%$
confidence level) with a flat de Sitter universe ($\4=1$ and $\5=-1$).
Given the large uncertainties, we cannot draw definitive conclusions
on this multi-band analysis. However, even if the data from the
$R$-galaxies have not information on the cosmology, the data from the
multiple arc and the last critical line in the $I$-band prefer
accelerating universes with subcritical matter density.

\section{Systematics}

In the previous section, we performed a statistical analysis based on
the data found in the literature. We want now to address some
systematics that can affect our results. A very accurate knowledge of
the mass distribution of the lens is required to put meaningful
constraints on cosmological parameters. One of the more important
source of indetermination comes from the modeling of the mass profile
of the lens (Chiba \& Takahashi \cite{ch+ta01}). In Eq.~(\ref{cri4}),
we have considered only the error coming from a not very accurate mass
normalization but, in general, we have also to face the
indetermination on the cluster mass profile. As a general feature, the
three-dimensional mass density of a clump, $\rho$, is proportional to
a typical length scale, $r_s$, so that, with respect to the angular
diameter distance, $\rho \propto D_d^{-2}$. The mass enclosed within
an angular radius $\theta$ comes out
\[
M(\theta ) \propto D_d \sigma^2 {\cal P} \theta,
\]
where $D_d$ contains the cosmological dependence and $\sigma^2$ stands
for an overall normalization. ${\cal P}$ is a factor accounting for
the deviations of the cluster mass profile from the SIS; ${\cal P}$ is
a function of $\theta$ and of some parameters, such as a core radius.
Substituting in Eq.~(\ref{cri2}), we get, for a spherically symmetric
lens,
\[
\frac{D_{ds}}{D_s} \propto \frac{\theta_t}{ \sigma^2 {\cal P} } .
\]
To consider the uncertainty in the profile, we have to add in
quadrature to the right hand side of Eq.~(\ref{cri4}) an additional
relative error of $\Delta {\cal P}/{\cal P}$. Usually, the main
contribution to the error budget comes from the mass normalization
but, in some extreme cases, the indetermination in the mass profile
can be of the same order. $\Delta {\cal P}$ is maximum when calculated
between a lens with a soft core and a halo with a singular density
steeply rising towards the center, as predicted by numerical
simulations in the standard cold dark matter framework of structure
formation and approximated by a Navarro-Frenk-White profile (NFW,
Navarro et al. (\cite{nav+al95,nav+al96})).

A NFW model can match the mass distribution of CL~0024+1654
(Broadhurst et al. \cite{bro+al00}). The required mass inside the
arc's radius for such a model, reproducing the projected mass
distribution outside the core radius, is $40\%$ higher than the
prediction of a power-law model (Tyson et al. \cite{tys+al98}).
Without an independent information, $\Delta {\cal P}$ would be a
really large error. Fortunately, the NFW profile is discarded since it
implies a velocity dispersion much higher than the measured value
(Shapiro \& Iliev \cite{sh&il00}). Once we can discard models with
singular central density, the indetermination in the model can be
accounted for by the errors within which we constrain the parameters
of the mass profile ($\Delta \gamma$ and $\Delta
\theta_c$ in the power-law model used in Sect.~3).
 In general, an uncertainty in the cluster mass profile can
significantly weaken the results on the cosmological parameters; but,
in the case of CL~0024+1654, the degeneracy in the fit is not the main
error.

Together with the overall mass profile, sub-structures must be
considered. In the case of a lens with a rather regular morphology,
even if a ``not correct" potential shape is used in the reconstruction
or the contribution of small sub-structures is neglected, the
cosmological parameters are still retrieved, although with larger
errors (Golse et al.~\cite{go&al01}). On the contrary, neglecting a
sub-structure as large as $20\%$ of the total mass in a bi-modal
cluster completely hides the effect of cosmology (Golse et
al.~\cite{go&al01}). Adding the contribution of individual galaxy
masses is also useful to tighten the confidence levels and can become
critical in some extreme cases, as a galaxy strongly perturbing the
location of multiple-images (Golse et al. \cite{go&al01}). Deep
imaging of CL~0024+1654 has made it possible to construct a
high-resolution map of the projected mass distribution of the cluster
and to take into account the effect of perturbing galaxies. Kassiola
et al.~(\cite{ka&al92}) and Wallington et al.~(\cite{wal+al95})
considered the perturbing potentials of two galaxies near the middle
segment of the arc. Tyson et al.~(\cite{tys+al98}) assigned one or
more mascons to each of the 118 cluster galaxies and 25 free mascons
for the remaining cluster mass. However, all these studies in
literature agree on a overall representation as the one in
Eq.~(\ref{cl1}).

Some features in the 3-D space, as a possible merger scenario, can
invalidate our estimation of the cosmological parameters. A recent
analysis of the distribution of the galaxies in the redshift space
(Czoske et al. \cite{czo+al01,czo+al02}) suggests a fairly complicated
structure. A group of galaxies lying just in front of the main cluster
could be the result of a high speed collision of two smaller clusters
with a merger axis very nearly parallel to the line of sight (Czoske
et al. \cite{czo+al02}). In particular, a bulk velocity component
present in the central velocity distribution would over-estimate the
mass obtained from the formal central velocity dispersion.
Furthermore, galaxies at large projected distance from the centre are
also affected by the collision and cannot be used to derive $\sigma$
(Czoske et al. \cite{czo+al02}). The consequences on the dark energy
constraint are quite dramatic since such a scenario could entail a
systematic error on the estimation of the velocity dispersion of the
same order of the statistical one. As can be seen from
Fig.~(\ref{omega_M-w_X_tys}), this additional error would completely
hide any dependencies on the cosmological parameters.

Even if, together with strong lensing observations, both a weak
lensing analysis out to $10$ arcmin (Bonnet et al. \cite{bon+al94})
and X-ray observations (Soucail et al. \cite{sou+al00}; B\"{o}hringer et
al. \cite{boh+al00}) favour a regular morphology, the points just
discussed suggest caution in the interpretation of the results
obtained in the previous section.

\section{Some final considerations}

We have explored the feasibility of reconstructing the properties of
the dark energy in the universe by using strong lensing systems in
which a cluster of galaxies acts as deflector. With respect to other
lensing systems, for the one discussed in this paper it is possible to
determine the position of the critical lines in two independent ways:
with giant arcs or with the radial shape of the depletion curves. This
circumstance allows to study  the ratio of angular diameter distances
that characterizes the angular position of the critical lines over a
large range of source redshifts, just for a single lensing cluster.
Provided that the properties of the background populations are well
constrained, it is possible in principle to use multi-band depletion
measurements to obtain several independent estimates of the ratio of
distances, each one probing a different source redshift.

For a flat universe, the sensitivity of the angular positions of the
critical lines on quintessence becomes higher in low-density
pressureless matter universes and for dark energy with intermediate
equation of state. While the analysis of only a few lensing clusters
suffices to distinguish between accelerating and decelerating models
of universe (also without a prior knowledge of $\1$), a considerably
larger sample ($N \sim 200$) and an accurate estimate of $\1$ are
needed to constrain the equation of state within an uncertainty of
$\Delta w_X \sim 0.25$ and discriminate, at the $95\%$ confidence
limit, between a cosmological constant and an evolving quintessence.

In our opinion, a first application of the method to the cluster CL
0024+1654 has given interesting results. A combined analysis of both
the multiple arc and the depletion curves disfavours models of a
universe without dark energy. On the other hand, flat accelerating
universes are in agreement with the data. These preliminary estimates
agree with the currently favoured constraints from other independent
measurements. However, some features in the redshift space of
CL~0024+1654, as a possible merger scenario, could invalidate our
results. Indeed, a very accurate knowledge of the absolute mass
distribution of the deflector and a correct understanding of the
pattern of sub-structures are necessary to obtain secure constraints
on the cosmological parameters.

The method we have discussed is quite general and can be applied to
several strong lensing systems. For example, a single galaxy, whose
stellar velocity dispersion can be accurately measured, can multiply
image a background quasar. Clusters of galaxies need an accurate
modeling of the pattern of substructures and present a quite
problematic measurement of $\sigma$ but allow one to study the ratio
$D_{ds}/D_s$ at different source redshifts. Furthermore, a multiple
image system of galaxies with known redshift makes possible an
absolute calibration of the total mass of the cluster (Golse et al.
\cite{go&al01}). We have shown how analyses of magnification bias in
multi-band photometry can be combined with observations of giant arcs
to obtain some insight on cosmological parameters.

\begin{acknowledgements}
I am very grateful to E. Piedipalumbo, C. Rubano, M.V. Sazhin and P.
Scudellaro for useful discussions and S. Refsdal for the reading of
the manuscript. I also thank the referee, J.-P. Kneib, for his reports
which helped to improve the paper.

\end{acknowledgements}

\end{document}